\documentstyle[12pt]{article}
\begin{document}

\begin{center}
Classical mechanics is not $\hbar\rightarrow 0$ limit of quantum mechanics\\
O.~V.~Man'ko, and V.~I.~Man'ko\\
Lebedev Physical Institute, Moscow, Leninskii pr., 53
\end{center}

\begin{abstract}
Both the set of quantum states and the set of classical states described by symplectic tomographic 
probability distributions (tomograms) are studied. It is shown that the sets have common part but 
there exist tomograms of classical states which are not admissible in quantum mechanics and vica versa, 
there exist tomograms of quantum states which are not admissible in classical mechanics. Role of different 
transformations of reference frames in phase space of classical and quantum systems (scaling and rotation) 
determining the admissibility of the tomograms as well as the role of quantum uncertainty relations is elucidated. 
Union of all admissible tomograms of both quantum and classical states is discussed in context of interaction of 
quantum and classical systems. Negative probabilities in classical mechanics and in quantum mechanics corresponding 
to the tomograms of classical states and quantum states are compared with properties of nonpositive and nonnegative 
density operators, respectively.
\end{abstract}

\section{Introduction}
The classical mechanics in quantum-like description was considered in~\cite{Olga}-\cite{VManko}. 
The possibility to combine description of quantum and classical degrees of freedom was discussed in~\cite{Shirokov}. 
In standard formulation of classical mechanics of systems with fluctuations (classical statistical mechanics) one 
uses the notion of the system state expressed in terms of probability distribution function $f(q,p)$ in the phase space 
of the system. Here $q$ is the position and $p$ is the momentum, which are supposed to be simultaneously measurable. 
In standard formulation of  quantum mechanics the notion of the system state is expressed in terms of complex wave function $\psi(q)$ 
which is interpreted as a "probability wave" (pure state). In case of mixed state it is described by density matrix which is a 
complex function of two variables $\rho(q,q')$ considered as matrix element $<q|\hat\rho|q'>$ of the nonnegative hermitian 
density operator $\hat\rho$ in position representation. Recently a new formulation of notion of state in classical mechanics and 
in quantum mechanics was suggested~\cite{Mancini1,Olga}. The state is described by positive measurable probability distribution function 
$w(X,\mu,\nu)$ (called tomogram) of random variable $X$ which has the physical meaning of the system position. 
This position $X$ is determined in the reference frame in the system phase space and the axis of the reference frame 
are scaled and rotated. The real parameters $\mu$ and $\nu$ label the reference frame in the phase space and the scaling 
parameters $\lambda$ and rotation angle $\theta$ are connected with the parameters $\mu$ and $\nu$ by the relation: 
$$\mu=e^{\lambda}\cos\theta\,\quad\quad \nu=e^{-\lambda}\sin\theta.$$
The position and momentum in quantum mechanics obey to uncertainty relation which prohibits simultaneous measuring 
these physical observables. The wave function, density matrix and tomogram of a quantum state depend on Planck constant 
$\hbar$. The Planck constant determines the uncertainty of the position and momentum in quantum mechanics. The less is  
the value of the Planck constant considered as a parameter, the less is the uncertainty of the position and momentum 
connected with the quantum fluctuations. It means that one can scale Planck constant $\hbar\rightarrow \epsilon\hbar$. 
For $\epsilon=-1$ this transform is connected with time reverse or with transposition of density matrix. The Planck 
constant is not involved into description of states in classical mechanics. 
There is a common vision of the connection of the quantum and classical mechanics. In this common vision the classical mechanics is the 
limit of the quantum mechanics for Planck constant equal to zero. It means that condition $\hbar=0$ reconstructs the simultaneous measurability 
of the position and momentum. The aim of our work is to discuss this problem. We will show that the picture of mutual 
relation of the quantum mechanics and classical mechanics differes from this  simple $\hbar\rightarrow 0$ limit picture. 
The reason is that the sets of states (sets of tomograms) in classical and quantum mechanics are different though 
they have the common part. In the $\hbar\rightarrow 0$ limit set of quantum states does not coincide with the set of 
classical states. This picture became possible to clarify namely in the tomographic probability representation~\cite{Mancini1}-\cite{MAMankoJRLR} 
for which both classical and quantum states are described by the same object (tomographic probability distribution). In representations 
using different objects (wave function and density matrix in quantum domain and probability distribution in phase space in classical 
domain) the comparison of situation in the limit $\hbar\rightarrow 0$ in the density matrix with classical situation is difficult.

\noindent Another goal of our paper is to construct formalism of classical mechanics using the operators in Hilbert space and their tomographic symbols like in quantum domain. The 
difference with quantum mechanics is connected with difference of product of the operators (star-product) used in classical domain which is 
commutative in this case. We will show that the notion of negative probabilities which usually is associated to the Wigner function in 
quantum mechanics can naturally appear in classical domain too. This follows from the fact that in quantum-like formalism of classical 
mechanics the density operators describing the classical states are hermitian but can be nonpositive. This means that in their spectral 
decomposition the eigenvalues (playing the role of probabilities) can take negative values. Thus one has interesting duality. In classical 
mechanics the probability distributions describing the classical states are nonnegative functions. But corresponding density 
operators describing the classical states in framework of quantum-like formalism can be nonpositive. In quantum domain the 
situation is opposite. The density operators describing the quantum states are nonnegative (eigenvalues in the spectral 
decomposition of the density operators playing the role of probabilities are mandatory positive). But the Wigner functions which play 
the role of probability density can take negative values in some domains of phase space. 

\section{Classical states and probability distribution in phase space}
First we discuss notion of classical states. For simplicity we consider a particle with one degree of freedom with mass 
$m=1$ which has position $-\infty<q<\infty$ and the velocity $-\infty<\dot q<\infty$. The state of the particle is 
completely determined by these two quantities $q$ and $\dot q$. We introduce the 
momentum $p=m\dot q$ which in our units ($m=1$) gives $p=\dot q$. Thus the state of the particle is identified with the 
point in phase space (plane) with coordinates $q,\, p$. Evolution of the particle state is described by a trajectory 
in the phase space $q(t),\, p(t)$. If one has the classical particle inside of some environment (interacting with 
other particles, e.g. in a gas with temperature $T$) the position $q$ and momentum $p$ fluctuate. In view of these 
fluctuations the state of the particle is described by a probability distribution function $f(q,p)$ which is nonnegative 
\begin{equation}\label{eq.1}
f(q,p)\geq 0 
\end{equation} 
and normalized
\begin{equation}\label{eq.2}
\int f(q,p)\,d q\,d p=1.  
\end{equation}
All the states of classical particles are described by the probability distribution functions belonging to a set. 
There are some other properties of the probability distribution functions from this set. These properties seem to 
be obvious according to our classical intuition. For example, we assume that to a state of the particle with 
velocity $\dot q$ (momentum $p$) and position $q$ corresponds another state with opposite velocity $-\dot q$ 
(momentum $-p$) and the same position $q$. In the case of states with fluctuations it means that if the function 
$f(q,p)$ belongs to the set of admissible probability distribution functions the function $f_t(q,p)=f(q,-p)$ also 
belongs to the set. It is the property of time reversilibity which we assume for the classical states. Analogously, 
the reflection in mirror operation for the position $q\rightarrow -q$ combined with unchanged momentum provides the 
following property. If a function $f(q,p)$ belongs to the set of admissible classical probability distributions the 
function $f_{-}(q,p)=f(-q,p)$ also belongs to this set. Combination of these two properties provides the result that 
if the function $f(q,p)$ belongs to the set of the probability distributions the function $f(-q,-p)$ also belongs to the 
same set. 
The possibility to shift the origin of the reference frame in the phase space of the particle which 
we accept intuitively as obvious one means that the probability distribution function $f(q+q_o,p+p_o)$ with arbitrary 
real shift parameters $q_o$ and $p_o$ also belongs to the set of admissible classical probability distributions. The 
discussed transforms belong to real symplectic group acting on classical phase space with addinbg the reflections.

\section{Scaling transform}
Now we will discuss the behaiviour of probability distribution function of a classical state with respect to independent 
scaling of position and momentum 
\begin{eqnarray}
&&q \rightarrow \lambda_q q\nonumber\\
&&p\rightarrow\lambda_p p,
\label{eq.S1}
\end{eqnarray}
where $\lambda_q$ and $\lambda_p$ are arbitrary real parameters. In fact, we consider reference frames in phase space of 
classical particle where the position and momentum axes are scaled. The transform~(\ref{eq.S1}) provides the change of 
probability distribution function 
\begin{equation}\label{eq.S2}
f(q,p)\rightarrow f_s(q,p)=N(\lambda_q,\lambda_p)f(\lambda_q q,\lambda_p p). 
\end{equation}
Normalisation constant $N(\lambda_q, \lambda_p)$ gives the equality 
\begin{equation}\label{eq.S3} 
\int f_s(q,p) dq\,dp=1, 
\end{equation}
and it reads 
\begin{equation}\label{eq.S4} 
N(\lambda_q,\lambda_p)=|\lambda_q\lambda_p|
\end{equation}
In integral form one has the relation 
\begin{equation}\label{eq.S5} 
f_s(q,p)=\int|\lambda_q\lambda_p|f(q',p')\delta(q'-\lambda_q q)\delta(p'-\lambda_p p)dq'\,dp', 
\end{equation}
where the kernel of the integral scaling tranform reads 
\begin{equation}\label{eq.S6}
K_s(q,p,q',p')=|\lambda_q\lambda_p|\delta(q'-\lambda_q q)\delta(p'-\lambda_p p). 
\end{equation}
The change of signs of $q$ and $p$ is the partial case of the transform~(\ref{eq.S5}) for 
$\lambda_q=-1,\, \lambda_p=-1$. In classical case if the distribution function $f(q,p)$ in phase space is admissible 
the functions $f_s(q,p)$~(\ref{eq.S5}) are also admissible for all real values of parameters $\lambda_q\neq 0$ and 
$\lambda_p\neq 0$. The real numbers form two-parameter commutative Lie group with product of group elements defined 
by standard multiplication rule of the real numbers. Geometrically the real numbers are determined by points on the 
plane $\lambda_q,\lambda_p$ with excluded points on the axes.  

\noindent If the variances and covariance of two fluctuating observables $q$ and $p$ in the state with the probability 
distribution density $f(q,p)$  are equal $\sigma_{q q}$, $\sigma_{p p}$ and $\sigma_{q p}$, respectively, the variances 
and covariance in the state with the probability distribution density $f_s(q,p)$~(\ref{eq.S5}) read 
\begin{equation}\label{eq.S7}
\sigma_{q q}^{(s)}=\lambda_q^{-2}\sigma_{q q},\,
\sigma_{p p}^{(s)}=\lambda_p^{-2}\sigma_{p p},\,
\sigma_{q p}^{(s)}=(\lambda_q\lambda_p)^{-1}\sigma_{q p}. 
\end{equation}
For symmetric dispersion matrix of the classical state with probability distribution $f(q,p)$, i.e., 
\begin{equation}\label{eq.S8}
\sigma=\pmatrix{\sigma_{q q}&\sigma_{q p}\cr\sigma_{q p}&\sigma_{p p}} 
\end{equation}
one can calculate the determinant 
\begin{equation}\label{eq.S9}
d=\sigma_{qq}\sigma_{pp}-\sigma_{q p}^2 
\end{equation}
and the trace 
\begin{equation}\label{eq.S10}
\mbox{T} =\sigma_{q q}+\sigma_{p p} 
\end{equation}
The dispersion matrix is nonnegative, i.e., 
\begin{equation}\label{eq.S11}
\sigma_{q q}\geq 0,\,\sigma_{p p}\geq 0,\, d\geq 0, \mbox{T}\geq 0. 
\end{equation}
For the classical state with the probability distribution $f_s(q,p)$ one has 
\begin{equation}\label{eq.S12} 
d^{(s)}=|\lambda_q \lambda_p|^{-2}d,\,\mbox{T}^{(s)}=\lambda_q^{-2}\sigma_{q q}+\lambda_p^{-2}\sigma_{p p}. 
\end{equation}
For all classical states of the particle the parameters $d^{(s)}$ and $\mbox{T}^{(s)}$ satisfy the 
inequalities~(\ref{eq.S11}). This property corresponds to the obvious statement that in classical case one 
can have states without fluctuations and correlations of position and  momentum as well as the states with 
fluctuations and correlations obeying only to constraints~(\ref{eq.S11}). 

\section{Tomographic representation} 
Let us discuss the property of classical states considered in previous sections using tomographic probability 
representation introduced in~\cite{Olga}. Folllowing this work we introduce the tomogram (called also 
tomographic probability distribution or marginal distribution) 
\begin{equation}\label{eq.T1} 
\omega(X,\mu,\nu)=\int f(q,p)\delta(X-\mu q-\nu p)d q\,d p, 
\end{equation}
where $\mu$ and $\nu$ are real parameters $-\infty<\mu,\,\nu<\infty$. One has normalisation condition 
\begin{equation}\label{eq.T2} 
\int\omega(X,\mu,\nu)\,d X=1
\end{equation}
for arbitrary parameters $\mu$ and $\nu$. Also one has the inverse formula 
\begin{equation}\label{eq.T3}
f(q,p)=\frac{1}{2\pi}\int\omega(X,\mu,\nu)e^{i(X-\mu q-\nu p)}d X\,d\mu\,d\nu. 
\end{equation}
The tomogram has the homogeneity property 
\begin{equation}\label{eq.T4}
\omega(\lambda X, \lambda\mu,\lambda\nu)=\frac{1}{|\lambda|}\omega(X,\mu,\nu).
\end{equation}
It is equal to probability density of the position $X$ measured in reference frame in phase space with rotated 
and scaled axes. The rotation angle $\theta$ and scaling parameter $e^{\lambda}$ are connected with parameters 
$\mu$ and $\nu$ as it is duscussed in the Introduction. The tomogram completely determines the classical state 
of the particle. The dispersions of the position and momentum can be calculated using the dispersion of the 
variable $X$ given by formula 
\begin{equation}\label{eq.T5}
\sigma_{X X}(\mu,\nu)=\int X^2\omega(X,\mu,\nu)d X-\large(\int X \omega(X,\mu,\nu)d X\large)^2. 
\end{equation}
Thus, one has 
\begin{equation}\label{eq.T6}
\sigma_{q q}=\sigma_{X X}(1,0),\,\sigma_{p p}=\sigma_{X X}(0,1). 
\end{equation}
The covariance of position and momentum has the form 
\begin{equation}\label{eq.T6}
\sigma_{q p}= 2\pi\int d X\frac{\partial^2 \omega(X, \mu, \nu)}{\partial\mu\partial\nu}\mid_{\begin{array}{c}
\mu=0\\ \nu=0\end{array}}       
\end{equation}
Let us discuss now the properties of the tomogram. The change $p\rightarrow -p$ in the probability distribution 
provides the change $\nu\rightarrow -\nu$ in the tomogram. The change $q\rightarrow -q$ provides the change 
$\mu\rightarrow -\mu$ in the tomogram. The scaling transform $q\rightarrow\lambda_q q,\, p\rightarrow\lambda_p p$ 
and $f(q,p)\rightarrow f_s(q,p)$ provides the transform of the tomogram 
\begin{equation}\label{eq.T8}
\omega(X,\mu,\nu)\rightarrow\omega_s(X,\mu,\nu)=\omega(X,\frac{\mu}{\lambda_q},\frac{\nu}{\lambda_p}). 
\end{equation}
The kernel of the above scaling transform in integral form reads 
\begin{eqnarray}
&&\omega(X,\mu,\nu)\rightarrow\omega_S(X,\mu,\nu)=\int\omega(X',\mu',\nu')K_S(X,\mu,\nu,X',\mu',\nu')d X',d\mu',d\nu',\nonumber \\
&&K_S(X,\mu,\nu,X',\mu',\nu')=\delta(X-X')\delta(\frac{\mu}{\lambda_q}-\mu')\delta(\frac{\nu}{\lambda_p}-\nu').\label{eq.T9}
\end{eqnarray}

\section{Uncertainty relations}
In quantum mechanics there exist some inequalities called uncertainty relations. The most important are 
uncertainty relations for position and momentum introduced by Heisenberg (see, for example~\cite{145}) and by Schrodinger and 
Robertson~\cite{SCH30,Rob2930}. Review 
of the uncertainty relations  is given in~\cite{Dod}-
\cite{DodVV}. Let us derive the uncertainty relations. We use the obvious 
inequality that for an arbitrary operator $\hat F$ one has 
\begin{equation}\label{eq.U1} 
<\hat F^{+}\hat F>\geq 0. 
\end{equation}
Here we consider the mean value either for pure state $|\psi>$,i.e., 
\begin{equation}\label{eq.U2} 
<\psi|\hat F^{+}\hat F|\psi>\geq 0,
\end{equation}
or for mixed state with nonnegative density operator $\hat\rho$, i.e.
\begin{equation}\label{eq.U3}
\mbox{Tr}(\hat\rho\hat F^{+}\hat F)\geq 0. 
\end{equation}
Let us construct a special operator $\hat F$ considering $N$ operators $\hat Q_{\alpha}$, where $\alpha=1,2,...,N,$ in the form 
\begin{equation}\label{eq.U4}
\hat F=\sum_{\alpha=1}^N C_{\alpha}\hat Q_{\alpha}.
\end{equation}
Here $C_{\alpha}$ are arbitrary complex numbers. Inequality~(\ref{eq.U1}) yields 
\begin{equation}\label{eq.U5}
\sum_{\alpha,\beta=1}^{N} C_{\alpha}^*<\hat Q^+_{\alpha}\hat Q_{\beta}>C_{\beta}\geq 0.
\end{equation}
Using the identity 
\begin{equation}\label{eq.U6}
\hat Q_{\alpha}^+\hat Q_{\beta}=\frac{1}{2}[\hat Q^+_{\alpha},\hat Q_{\beta}]+\frac{1}{2}\{\hat Q^+_{\alpha},\hat Q_{\beta}\} 
\end{equation}
and introducing the notation 
\begin{equation}\label{eq.U7} 
\frac{1}{2}<\{\hat Q^+_{\alpha},\hat Q_{\beta}\}>=\sigma_{\alpha\beta}, 
\end{equation}
\begin{equation}\label{eq.U8}
\frac{1}{2}<[\hat Q^+_{\alpha},\hat Q_{\beta}]>=\Sigma_{\alpha\beta}, 
\end{equation}
one has inequality~(\ref{eq.U5})  in  the form 
\begin{equation}\label{eq.U9} 
\sum_{\alpha\beta=1}^N C_{\alpha}^*(\sigma_{\alpha\beta}+\Sigma_{\alpha\beta})C_{\beta}\geq 0. 
\end{equation}
The nonnegativity condition for this quadratic form implies the hermitian matrix of this form satisfies the positivity 
condition, i.e.,  
\begin{equation}\label{eq.U10}
(\sigma_{\alpha\beta}+\Sigma_{\alpha\beta})\geq 0. 
\end{equation}
The derived inequality is general one. Let us consider important partial cases. For one degree of freedom $(\alpha=1,2)$ 
let us take hermitian operators 
\begin{equation}\label{eq.U11}
\hat Q_1=\hat q-<\hat q>,\quad \hat Q_2=\hat p-<\hat p>.
\end{equation}
In this case the matrix $\sigma_{\alpha\beta}$ is the dispersion matrix for position and momentum. 
The matrix $\Sigma_{\alpha\beta}$ has the form proportional to Pauli matrix $\sigma_y$, i.e. 
\begin{equation}\label{eq.U12} 
\Sigma_{\alpha\beta}=\pmatrix{0&\frac{i}{2}\cr-\frac{i}{2}&o} .
\end{equation}
(We take below $\hbar=1)$.

\noindent The inequality~(\ref{eq.U10}) means that the matrix 
\begin{equation}\label{eq.U13}
\pmatrix{\sigma_{q q}&\sigma_{q p}+\frac{i}{2}\cr\sigma_{q p}-\frac{i}{2}&\sigma_{p p}}\geq 0. 
\end{equation}
The criterion of positivity of the matrix means 
\begin{equation}\label{eq.U14} 
\sigma_{q q}\geq 0,\quad \sigma_{p p}\geq 0 
\end{equation} 
and
\begin{equation}\label{eq.U15} 
\sigma_{q q}\sigma_{p p}-\sigma^2_{q p}-\frac{1}{4}\geq 0. 
\end{equation} 
Inequality~(\ref{eq.U15}) is the Schr\"odinger--Robertson uncertainty relation~\cite{Dod} 
\begin{equation}\label{eq.U16} 
\sigma_{q q}\sigma_{p p}\geq \frac{1}{4(1-r^2)}. 
\end{equation}
Here $r$ defines the correlation of position and momentum 
\begin{equation}\label{eq.U17}
r=\frac{1}{\sqrt{\sigma_{q q}\sigma_{p p}}}\sigma_{q p}. 
\end{equation} 
For $n$ degrees of freedom let us take $N=2n$ operators $\hat Q_{\alpha}$ in the form 
\begin{eqnarray}  
&&\hat Q_1=\hat q_1-<\hat q_1>,\quad\hat Q_2=\hat q_2-<\hat q_2>,\quad ...\,,\quad 
\hat Q_n=\hat q_n-<\hat q_n>,\nonumber\\
&&\hat Q_{n+1}=\hat p_1-<\hat p_1>,\quad 
\hat Q_{n+2}=\hat p_2-<\hat p_2>,\quad ...\,,\quad
\hat Q_N=\hat p_n-<\hat p_n>. \label{eq.U18}
\end{eqnarray} 
In this case the Robertson uncertainty relation~\cite{145} expressed as the condition of 
positivity of $2n\times2n$- matrix~(\ref{eq.U10}), where $\sigma_{\alpha\beta}$ is multimode dispersion 
matrix of positions ans momenta and the $2n\times2n$-matrix $\Sigma_{\alpha\beta}$ has the form~(\ref{eq.U12}) 
with $n\times n$ - blocks proportional to identity matrix in $n$ dimensions. 

The positivity condition~(\ref{eq.U10}) means that all the major minors of the $2n\times2n$-matrix are nonnegative. 
The determinant of the matrix~(\ref{eq.U10}) can be calculated and one has the uncertainty relation in the 
form of the inequality 
\begin{equation}\label{eq.U19}
\mbox{det}(\sigma_{\alpha\beta})\geq \frac{1}{4^n}. 
\end{equation} 
One can take the operator $\hat Q_{\alpha}$ as angular momentum operators (spins). In this case one has 
inequalities which give some constraints for spin states. 

\section{Admissibility of quantum states} 
In the case  of  classical statistical system with $n$ degrees of freedom one can determine the dispersion 
matrix of position and momentum 
\begin{equation}\label{eq.U20} 
\sigma_{\alpha\beta}=\int f(\vec q,\vec p) Q_{\alpha}Q_{\beta}d\vec q d\vec p. 
\end{equation}
Here $f(\vec q,\vec p)$ is probability density in the phase space of the system, 
$$\int f(\vec q,\vec p)d \vec q d\vec p=1.$$ 
The functions 
$Q_{\alpha},\, \alpha=1,2,\,...\,,N=2n$   
are defined as 
\begin{eqnarray}  
&& Q_1= q_1-<q_1>,\quad Q_2= q_2-<q_2>,\quad ...\,,\quad
 Q_n= q_n-< q_n>,\nonumber\\
&& Q_{n+1}= p_1-< p_1>,\quad
Q_{n+2}= p_2-< p_2>,\quad ...\,,\quad
Q_N= p_n-< p_n>. \label{eq.U21}
\end{eqnarray} 
and 
\begin{eqnarray}
&&<q_s>=\int f(\vec q,\vec p)q_s d\vec q d\vec p,\nonumber\\
&&<p_s>=\int f(\vec q,\vec p)p_s d\vec q d\vec p,\nonumber\\ 
&&s=1,2,\, ...\,,\,n.\label{eq.U22}
\end{eqnarray} 
Intuitively it is clear that one can make scaling transform of the probability density 
\begin{equation}\label{eq.U23}
f(\vec q,\vec p)\rightarrow f_s(\vec q,\vec p)=\int K(\vec q,\vec p,\vec q\,',\vec p\,')f(\vec q\,',\vec p\,') 
d \vec q\,' d \vec p\,', 
\end{equation} 
where the kernel has the form 
\begin{equation}\label{eq.U24}  
K(\vec q,\vec p,\vec q\,',\vec p\,')=\prod_{s=1}^n |\lambda_{q_s}\lambda_{p_s}|
\delta(q_s'-\lambda_{q_s}q_s)\delta(p_s'-\lambda_{p_s}p_s). 
\end{equation} 
The $2n$ real parameters $\lambda_{q_s},\,\lambda_{p_s}$ are arbitrary nonzero numbers. These parameters can be 
considered as abelian group which is direct product on n abelian groups discussed for one degree of freedom. 
The distribution~(\ref{eq.U23}) is admissible to describe a classical state. 
Quantum states with density operator $\hat\rho$ can be described by the Wigner function~\cite{Wig32} 
$W(\vec q,\vec p)$.  This function is connected with the density matrix in position representation $\rho(\vec x,\vec x\,')$
by the formulae 
\begin{eqnarray}
&&W(\vec q, \vec p)=\int\rho(\vec q+\frac{\vec u}{2},\vec q -\frac{\vec u}{2})e^{-i\vec p\vec u}d\vec u, \nonumber\\
&&\rho(\vec x,\vec x\,')=\frac{1}{(2\pi)^n}\int W\left(\frac{\vec x+\vec x\,'}{2}, \vec p\right)
e^{i\vec p(\vec x-\vec x\,')}d\vec p. \label{eq.Q1}
\end{eqnarray} 
The mean values and dispersion matrix for positions and momenta of quantum system can be calculated using 
formulas~(\ref{eq.U22}) and~(\ref{eq.U20}) with replacements 
\begin{equation}\label{eq.Q2} 
f(\vec q,\vec p)\rightarrow \frac{W(\vec q,\vec p)}{(2\pi)^n}. 
\end{equation}
The Wigner function describing the quantum state must satisfy the uncertainty relations. Due 
to this if one has scaling transform 
\begin{equation}\label{eq.Q3}
W(\vec q,\vec p)\rightarrow W_s(\vec q,\vec p)=\int K(\vec q,\vec p,\vec q\,',\vec p\,')
W(\vec q\,',\vec p\,')d\vec q\,'d\vec p\,' 
\end{equation}
the kernel of scaling transform is given by~(\ref{eq.U24}) but the parameters $\lambda_{p_s}$ and $\lambda_{q_s}$ must 
satisfy constraints providing the uncertainty relations 
$$|\lambda_{p_s}\lambda_{q_s}|\geq 1.$$
It means that in comparison with classical states where the parameters form the commutative group in quantum state the 
parameters form the semigroup of real numbers which is direct product of semigroups corresponding to one degrees of 
freedom. The numbers $\lambda_{p_s},\,\lambda_{q_s}$ are situated on the plane of these parameters. For classical states 
the forbidden numbers $\lambda_{p_s},\,\lambda_{q_s}$ are situated on the "classical cross" which is composed by two 
axes of reference frame in the plane. The thickness of the cross components is zero. In quantum case one has 
"quantum cross" where the values of parameters $\lambda_{p_s},\,\lambda_{q_s}$ are forbidden by the uncertainty relations. 
The boundary of the cross is formed by four hyperbolas. 
Outside of the "quantum cross" one has admissible values of parameters $\lambda_{p_s},\,\lambda_{q_s}$  and all points of 
the plane in the admissible domain form the semigroup of numbers. Thus transition from classical domain to quantum domain 
can be associated with operation of reducing the group of real numbers with natural product formula to semigroup of these 
numbers (outside of quantum cross) with the same natural product formula. The thickness of the quantum cross is nonzero 
and it is proportional to Planck constant. The classcial states in classical statistical mechanics have the property that 
scaling of any subsystem is permitted. In quantum domain one has some states which behave differently with respect to 
scaling of some subsystem due to specific quantum correlations.

\noindent If one uses the tomographic representation of quantum states the states are associated with symplectic tomogram 
\begin{equation}\label{eq.K1}
W(\vec X, \vec\mu,\vec\nu)=
\int W(\vec q,\vec p)\left(\prod_{s=1}^n\delta(X_s-\mu_s q_s-\nu_s p_s)\right)\frac{d\vec q d\vec p}{(2\pi)^n}.  
\end{equation}
The tomogram is probability distribution function of position $\vec X$ depending on extra real parameters $\vec\mu$ and 
$\vec\nu$. The scaling transform of the tomogram reads 
\begin{eqnarray}
\omega(\vec X,\vec \mu,\vec\nu)\rightarrow\omega_s(\vec X,\vec\mu,\vec\nu)&=&\int W(\vec X\,',\vec\mu\,',\vec\nu\,')
\prod_{s=1}^n\delta(X_s-X_s')\nonumber\\
\times\delta(\mu_s'-\frac{\mu_s}{\lambda_{q_s}})\delta(\nu_s'-\frac{\nu_s}{\lambda_{p_s}})d\vec\mu\,' 
d\vec\nu\,'d\vec X\,'.\label{eq.K2}
\end{eqnarray}
The tomogram of the quantum states are admissible if the values of parameters yield the uncertainty relation. 
It means that the linear map of the density operator $\hat\rho$ of the quantum state with the tomogram 
$\omega(\vec X,\vec\mu,\vec\nu)$~\cite{mmsz} is positive map. The states can be devided into two classes. 
Some states are admissible if cooordinates of any subsystem $\mu_s,\,\nu_s$ are scaled. In other cases 
scaling of subsystem coordinates $\mu_s,\,\nu_s$ provides the nonpositive density operator.

\section{Kernel of commutative star-product of density matrices of classical states}

In quantum mechanics observables and states are described by hermitian operators. The states are described by nonnegative 
density operators $\hat\rho$. The Weyl symbol $W(q,p)$ of arbitrary operator $\hat A$ is given in terms of the matrix 
of the operator $A(x,x')=<x|\hat A|x'>$ in position representation as 
\begin{equation}\label{eq.C1} 
W_A(q,p)=\int A(q+\frac{u}{2},q-\frac{u}{2})e^{-i p u} d u. 
\end{equation} 
The inverse of this formula reads 
\begin{equation}\label{eq.C2}
A(x,x')=\frac{1}{2\pi}\int W_A\left(\frac{x+x'}{2},p\right)e^{ip(x-x')}d p. 
\end{equation}
The density operator $\hat\rho$ is nonnegative hermitian operator which  means that 
\begin{equation}\label{eq.C3} 
\mbox{Tr}(\hat\rho|\psi><\psi|)\geq 0
\end{equation} 
for arbitrary projector $|\psi><\psi|$. The Wigner function $W(q,p)$ of the quantum state must satisfy the condition 
\begin{equation}\label{eq.C4}
\int\psi^*(x)W\left(\frac{x+x'}{2},p\right)e^{ip(x-x')}\psi(x')d p\, d x\, d x'\geq 0 
\end{equation} 
for arbitrary wave function $\psi(x)$. The positivity condition means that the quantum uncertainty relation calculated by 
means of the given Wigner function is necessary condition for admissible quantum state. 

\noindent For classical state the positive probability distributions $f(q,p)$ (states) and the functions in phase space 
$A(q,p)$ (observables) can be considered as Weyl symbols of some hermitian operators $\hat\rho_{cl}$ and $\hat A_{cl}$,  
respectively. Since the uncertainty relation can be violated in classical states it means that the corresponding 
density operators $\hat\rho_{cl}$ of classical state can be considered as nonpositive operator. In the spectral 
decomposition of the classical density operator 
\begin{equation}\label{eq.C5}
\hat\rho_{cl}=\int W_{cl}(j)|\psi_j><\psi_j|d j,  
\end{equation}
the weight function $W_{cl}(j)$ playing the role of probability density can take negative values which is mandatory 
for the classical state with density $f(q,p)$ violating quantum uncertainty relation. Thus we have the property. Positive 
probability density of the classical state $f(q,p)$ being expressed in terms of density operator $\hat\rho_{cl}$ has real quantum-like 
probability density $W_{cl}(j)$ which can take negative values. This property is dual to the property of Wigner function. For quantum 
state $\hat\rho$ (nonnegative density operator) the Wigner function $W(q,p)$ playing the role of classical-like probability density 
in phase space can take negative values (negative probability). Thus, the "negative probabilities" can be associated both to classical 
and to quantum states. Classical states with positive densities $f(q,p)$ in quantum-like description by means of nonpositive density operators 
are associated with "negative probabilities" $W_{cl}(j)$ in spectral decomposition of the density operators violating the uncertainty relation. 
Quantum state have only positive probabilities associated with spectral decomposition of nonnegative density operators but 
they demonstrate "negative probabilities" which are Wigner functions associated with the quantum states. The Weyl symbols of two quantum 
observables $W_A(q,p) $ and $W_B(q,p)$ are multiplied using the kernel 
\begin{equation}\label{eq.C6}
W_A\star W_B(q,p)=\int W_A(q_1,p_1)W_B(q_2,p_2)K(q_1,p_1,q_2,p_2,q,p,)d q_1 d q_2 d p_1 d p_2. 
\end{equation}
The product of Weyl symbols of quantum observables is noncommutative since, see, e.g.~\cite{33} 
\begin{equation}\label{eq.C7}
K(q_1,p_1,q_2,p_2,q,p,)= \frac{1}{\pi^2}\exp\{2i[q_1(p_2-p_3)+q_2(p_3-p_1)+q_3(p_1-p_2)]\}.
\end{equation}
If one consideres the product of matrices of operators $\hat A$ and $\hat B$ in position representation $A(x,x')$ and $B(x,x')$ corresponding 
to classical observable $A(q,p)$ and $B(q,p)$ with commutative pointwise product rule the kernel of the star-product 
\begin{equation}\label{eq.C8}
A\star B(x,x')=\int A(x_1,x_2)B(x_1',x_2')K(x_1,x_2,x_1',x_2',x,x') d x_1 d x_2 d x_1' d x_2' 
\end{equation}
has the form 
\begin{eqnarray}
K(x_1,x_2,x_1',x_2',x,x')=&&\int d u_1 d u_2\delta(x-x'-u_1-u_2)\delta(x_1-\frac{x+x'}{2}-\frac{u_1}{2})\nonumber\\
&&\delta(x_2-\frac{x+x'}{2}+\frac{u_1}{2})
\delta(x_1'-\frac{x+x'}{2}-\frac{u_2}{2})\delta(x_2'-\frac{x+x'}{2}+\frac{u_2}{2}). \label{eq.C9}
\end{eqnarray}
Thus we obtain the kernel of commutative star-product of matrices of operators $\hat A$ and $\hat B$ in position representation for classical 
observables $A(q,p)$ and $B(q,p)$ with pointwise product.

\section{Quantum and classical subsystems}
In this section we discuss a possibility to describe interaction of classical and quantum systems. For this we start from the state of two 
independent systems. First system is considered as quantum one and the second system is considered as classical one. The tomographic description 
of the combined system can be given in terms of tomogram $\omega(X_1,\mu_1,\nu_1,X_2,\mu_2,\nu_2)$. If there is no correlations of 
quantum $(X_1)$ and classical $(X_2)$ degrees of freedom the tomogram has the factorized form 
\begin{equation}\label{eq.F1}
\omega(X_1,\mu_1,\nu_1,X_2,\mu_2,\nu_2)=\omega_q(X_1,\mu_1,\nu_1)\omega_{cl}(X_2,\mu_2,\nu_2). 
\end{equation} 
The tomogram $\omega_q$ satisfies the uncertainty relation. The classical tomogram $\omega_{cl}$ has no such constraints. 
The form~(\ref{eq.F1}) corresponds to factorized form of the density operator of combined system 
\begin{equation}\label{eq.F2}
\hat\rho=\hat\rho_q\hat\rho_{cl}.  
\end{equation}
One can construct some evolution equation for the tomogram containing the classical and the quantum parts. This problem 
needs further investigation. 
Another problem is $\hbar\rightarrow 0$ limit of quantum mechanics. Since all the density operators of quantum states are 
nonnegative and the density operators of classical states (in quantum-like description) can be nonpositive there is no 
possibility in the limit $\hbar\rightarrow 0$ to get from nonnegative operator the nonpositive ones. In framework of tomograms 
the quantum tomograms in classical limits give the result which can be not admissible. On the other hand there are 
classical tomograms which are not admissible in quantum mechanics. Due to this these tomograms can not be obtained by the limit 
procedure $\hbar\rightarrow 0$ from the quantum tomogram. This discussion can be elucidated also in terms of transition 
group~$\rightarrow$~semigroup. As we clarified the uncertainty relation in quantum mechanics implied the change of scaling 
group by scaling semigroup. This operation shows that we change the mathematical nature of object (scaling tranform) 
considering transition from classical to quantum domain. To conclude one can consider union of admissible tomograms as 
probability description of combined classical-quantum system. This kind of speculation gives a possibility in natural 
way to describe interaction of quantum and classical degrees of freedom. 

\section{Conclusion}
We summarize the main points of our consideration. 

\noindent 1. In classical statistical mechanics the states can be associated in usual representation with probability 
distribution function $f(q,p)$ in phase space of the system. In quantum-like representation the states in classical 
statistical mechanics can be associated with density operators $\hat\rho_{cl}$ acting in a Hilbert space. In tomographic 
representation the states in classical statistical mechanics can be associated with probability distribution function  
$\omega(X,\mu,\nu)$ of position $X$ measured in ensemble of reference frames labelled by real parameters $\mu$ and 
$\nu$. The density operators $\hat\rho_{cl}$ can be nonpositive. The distribution functions $f(q,p)$ and tomograms 
$\omega(X,\mu,\nu)$ are nonnegative functions.  

\noindent The observables in classical statistical mechanics are associated 
with functions in phase space A(q,p) in usual representation. In quantum-like representation the observables in classical 
statistical mechanics are described by hermitian operators $\hat A_{cl}$. The density operators $\hat\rho_{cl}$ and the 
observables in classical statistical mechanics can be given in position representation by the matrices 
$\rho_{cl}(x,x')=<x|\hat\rho_{cl}|x'>,\,A_{cl}(x,x')=<x|\hat A_{cl}|x'>$. For example, in classical statistical mechanics 
the position and momentum operators have the form $\hat q_{cl}=x$ and $\hat p_{cl}=-i\frac{\partial}{\partial x}$ or in matrix 
representation $q(x,x')=<x|\hat q_{cl}|x'>=x\delta(x-x')$, $p_{cl}(x,x')=<x|\hat p_{cl}|x'>=-i\delta'(x-x')$. The 
commutative product of the operator matrices is determined by means of the kernel given by~(\ref {eq.C9}). The classical evolution 
equation for classical density operators does not preserve the positivity of the initial density operator. This preservation 
of positivity takes place only for linear systems. In tomographic representation the observables in classical statistical mechanics 
are described by their tomograms $A(X,\mu,\nu)$. The commutative product of the tomograms corresponds to pointwise product of the 
observables $A(q,p)$ in usual representation and to commutative product of the operators $\hat A_{cl}$ in quantum-like 
representation. The tomograms and density operators of states in classical  statistical mechanics must satisfy the condition of 
positivity of the probability distributions $f(q,p)$ describing the states in usual representation. But uncertainty relations of 
quantum mechanics can be violated.

\noindent II. In quantum mechanics the states are described by nonegative density operators $\hat\rho$ in usual representation. In 
phase-space representation the states are associated with Wigner function $W(q,p)$ which can take negative values. In tomographic 
representation the quantum states are described by the tomogram which is nonegative probability distribution function of the 
position $X$ measured in ensemble of reference frames in phase space labelled by real parameters $\mu$ and $\nu$. The density operator 
must be nonegative. The observables can be described by their tomograms which correspond to standard operators describing the observable 
in usual representation of quantum mechanics. In phase space representation the observables are described by the functions $A(q,p)$ 
which are Weyl symbols of the operators $\hat A$. The product of these functions is noncommutative and it is described by known kernel. 
The uncertainty relation provides constraints on admissible tomograms and Wigner functions of the quantum states. These constraints are 
associated with commutative semigroup of scaling transforms of tomograms ( and Wigner functions ) in quantum mechanics. 
In classical statistical mechanics the admissible tomograms ( and Wigner functions ) of the states are related by commutative group of 
scaling transforms. The quantum evolution equations preserve the nonnegativity of the quantum density operators. This 
corresponds to respecting the uncertainty relations in the process of quantum evolution. There exists a possibility to consider a union 
of all tomograms admissible both in quantum mechanics and in classcial statistical mechanics to describe the interaction of classical 
and quantum systems. In the standard limit procedure $\hbar\rightarrow 0$ the set of quantum states (quantum tomograms) 
does not coincide with the set of classical tomograms.  

\noindent After the work was finished authors became aware of the work~\cite{Br} where analogous aspects of connection of Heisenberg 
uncertainty relation with nonnegativity of classical density operator were discussed. 

\section{Acknowledgements}
O.V.M. is grateful to the Russian Foundation for Basic Research for partial support of the work under Project No. 03-02-16408.

\end{document}